\documentstyle[prl,aps,multicol,epsfig]{revtex}
\def\reff#1{(\ref{#1})}
\def\dfrac#1#2{{\displaystyle{#1\over#2}}}

\begin{document}
\title{Superconducting pairing of spin polarons in the $t$-$J$ model}
\author{N.M. Plakida and V.S. Oudovenko}
\address{
Joint Institute for Nuclear Research, 141980 Dubna, Russia}
\author{P. Horsch and A. I. Liechtenstein}
\address{Max Planck Institut f\"{u}r Festk\"{o}rperforschung,
Heisenbergstr. 1, D-70569 Stuttgart, Germany.}

\date{\today}
\maketitle

\begin{abstract}
The pairing of quasiparticles  in a CuO$_2$ plane is studied within
a spin polaron formulation of the $t$-$t^{'}$-$J$ model.
%
%
%
%
%
Our numerical solution of the Eliashberg equations
unambiguously shows $d$-wave pairing
between  spin polarons on different sublattices
mediated by the exchange of spin-fluctuations,
and a strong doping dependence of the quasiparticle bandwidth.
The transition temperature $T_c$ is an increasing
function of $J/t$ and crosses a maximum at an optimal doping
concentration $\delta_{opt}$.
For the $t$-$J$ model with $J/t=0.4$ we obtain $T_c\simeq 0.013 t$ at
$\delta_{opt}\simeq 0.2$.

\end{abstract}


\begin{multicols}{2}


\narrowtext
Recent experimental evidence in favor of a $d$-wave
superconducting pairing in high-$T_{c}$ cuprates ~\cite{1a}
supports theoretical studies of models with
strong electron correlations ~\cite{1b}.
The minimal model describing hole motion in CuO$_2$ plane
is the $t$-$J$~\cite{1d} or $t$-$t'$-$J$~\cite{1e}   model.
Numerical studies~\cite{1b,1ba,1c}  of small $t$-$J$ clusters
suggest a $d$-wave superconducting instability.
Yet to elucidate the nature of this
pairing, an analytical treatment of the $t-J$ model is
needed. For this purpose we use a spin polaron
formulation for the $t-J$  model deduced in the
region  of small hole concentrations. A
number of studies of this model~\cite{5,6}
predict that doped holes dressed by strong
antiferromagnetic spin fluctuations propagate
coherently as quasiparticles (spin polarons) with weight
$Z_{k}\simeq J/t$. It is quite natural to
expect that the same spin fluctuations induce
superconducting pairing of the spin polarons.  Recently
this problem  has been treated in the framework of the standard
BCS formalism assuming a rigid band model for the quasiparticles
~\cite{8a,8b}.  However, since the
spin-fluctuation energy is of the same order as a
quasiparticle bandwidth $\sim J$ a strong coupling approach is
necessary.

In this paper we present the first consistent solution of
the strong coupling spin polaron model at finite
temperatures and hole concentrations for normal and
superconducting states.  A numerical solution
of a self-consistent system of equations for hole and magnon Green
functions proves singlet $d$-wave superconducting pairing.
The gap function shows interesting additional structure on top
of the simple $\Delta_{k} = \Delta_0 (\cos{k_x}-cos{k_y})$ which reflects
the Fermi surface geometry. The doping dependence of $T_c$ around
$\delta_{opt}$ has the form of an inverted parabola, similar to
experiment, and a $T_c^{max}\sim 60 K$.
Combining these results with already existing weak coupling studies
for the Hubbard model~\cite{10,13}
we argue that  the spin-exchange pairing is
the true mechanism for high-temperature superconductivity
as proposed earlier by several groups based on more
phenomenological approaches~\cite{1a,14,15,16}.

We will study a spin polaron model on a two sublattice antiferromagnetic
(AF) background which has been successfully tested in the single hole
case ~\cite{5},~\cite{6}.
Spinless fermion operators $h_i^+$ and $f_i^+$ are introduced for holes on
different sublattices, i.e. on the $\uparrow(\downarrow)$-sublattice
the constrained electron operators $\tilde
c_{i\sigma}=c_{i\sigma}(1-n_{i -\sigma})$ of the $t$-$J$ model are
replaced by $\tilde c_{i \uparrow} = h_{i}^{+}, \;
\tilde c_{i \downarrow} = h_{i}^{+} S_{i}^{+} $
$ (\tilde c_{i \downarrow} = f_{i}^{+}, \;
\tilde c_{i \uparrow} = f_{i}^{+} S_{i}^{-}) $,
where $S_{i}^{\pm}=S_{i}^{x}\pm S_{i}^{y}$ are  spin operators.
This representation excludes doubly occupied states  and takes
into account strong AF spin correlations in the electron hopping.

By employing the linear spin-wave approximation
 in terms of the Holstein-Primakoff operators:
 $S_{i}^{+} \simeq  a_{i} , \;  (i \in \uparrow) ,  \; \; \;
 S_{i}^{+} \simeq  b_{i}^{+} , \;(i \in \downarrow) $
and performing  the Bogoliubov canonical transformation:
 $a_{k}= v_{k} \alpha_{k} + u_{k} \beta_{- k}^{+} , \; \;
 b_{k}= v_{k} \beta_{k} + u_{k} \alpha_{- k}^{+} ,$
 we obtain the spin polaron model:
\begin{eqnarray}
&H&_{t-J}=\sum_{kq}^{}(h_{k}^{+}f_{k-q}[g(k,q)\alpha_{q}+
g(q-k,q)\beta^{+}_{-q}]+\mbox{h.c.}) \nonumber\\
&+&\sum_{k}\epsilon_{k}(h_{k}^{+}h_{k}+f_{k}^{+}f_{k})
+\sum_{q}^{}\omega_{q}(\alpha_{q}^{+}\alpha_{q}+\beta_{q}^{+}\beta_{q}).
\label{2}
\end{eqnarray}
Here
$g(k,q)= ({zt}/{\sqrt {N/2}}) (u_{q}\gamma_{k-q}+v_{q}\gamma_{k})$
is the hole-magnon interaction,
$z=4$ is the number of the nearest neighbors on a square lattice with
$N$ sites,
$u_{k}= \left( {(1+\nu_{k})}/{2\nu_{k}} \right)^{1/2}\; ,$
$v_{k}=-{\rm sign}(\gamma_{k})\left({(1-\nu_{k})}/{2\nu_{k}}
\right)^{1/2} $, $ \nu_{k}= \sqrt{1-\gamma_{k}^{2}},$
 $\gamma_{k}=\frac{1}{2}(\cos k_{x}+\cos k_{y})$.
The next nearest neighbor hopping energy is
$\epsilon_{k} = (4 t' \cos k_{x} \cos k_{y}-\mu) $.
The  chemical potential $\mu$  should be calculated
self-consistently as a function of a
hole concentration $\delta$ and temperature $T$ from the equation:
$\delta= \langle  h_{i}^{+}h_{i}\rangle  +
 \langle  f_{i}^{+}f_{i}\rangle$.
The spin-wave energy  is $\omega_{q}=SzJ(1-\delta)^{2} \nu_{q}$
where $(1-\delta)^{2}$ is the mean field renormalization factor.
We neglect here the contact hole-hole
interaction which is unimportant in the polaron pairing ~\cite{8b}.
The summation over wave-vectors  in \reff{2} and below is
restricted to  $N/2$ points in the AF Brillouin zone.

To discuss singlet superconducting pairing within the spin polaron model
\reff{2},
we consider the matrix Green function (GF) for holes on two sublattices
$G_{hh}(k,z)=
\langle\langle h_{k}^{+}\mid h_{k}\rangle\rangle _{z} =
\langle\langle f_{k}^{+}\mid f_{k}\rangle\rangle _{z}$ and the
anomalous GF
$G_{hf}(k,z)=
\langle\langle h_{k}^{+} \mid f_{-k}^{+}\rangle\rangle _{z} =
- \langle\langle f_{-k}^{+}\mid h_{k}^{+}\rangle\rangle _{z}$,
where Zubarev's notation ~\cite{17}
for the anticommutator GF
was used  with $z=\omega+i\epsilon$.
To obtain  self--consistent equations for these GF's
we  employ the self--consistent Born approximation (SCBA)
which provided good results for the one--hole spectrum in the
normal state~\cite{5,6,7}.
In SCBA we get for the self-energies
\begin{eqnarray}
\Sigma_{hh}(k,i\omega_{n})=-T\sum_{q,m}
G_{hh}(q,i\omega_{m})\lambda^{11}_{k,k-q}(\omega_{n}-\omega_{m}),\\
\label{4a}
\Sigma_{hf}(k,i\omega_{n})=-T\sum_{q,m}
G_{hf}(q,i\omega_{m})\lambda^{12}_{k,k-q}(\omega_{n}-\omega_{m}) .
\label{4b}
\end{eqnarray}
where the Matsubara frequencies $\omega_{n}=\pi T(2n+1)$.
The interaction functions are
\begin{eqnarray}
\lambda^{11}_{k,q}(\omega_{\nu})=
g^{2}(k,q) D(q,-i\omega_{\nu})+
g^{2}(q-k,q) D(-q,i\omega_{\nu}),\nonumber \\
\lambda^{12}_{k,q}(\omega_{\nu})=
g(k,q)g(q-k,q)\{ D(q,-i\omega_{\nu})+
 D(-q,i\omega_{\nu})\}.\nonumber
\end{eqnarray}
The diagonal magnon GF $D(q,\omega)=
\langle \langle \alpha_{q}\mid \alpha_{q}^{+}\rangle \rangle _{\omega}$
in the zero order approximation is given by $D^{0}(q,\omega)=
(\omega-\omega_{q})^{-1}$ with the doping dependent magnon energy
$\omega_{q}$.
The full magnon GF is determined by the matrix equation
$\hat D^{-1}(q,\omega)= (\hat D^{0})^{-1}(q,\omega) - \hat \Pi(q,\omega)$
where the  renormalization of the magnon energy due to particle-hole
excitations is described by the polarization operator  $\hat
\Pi(q,\omega)$. This is calculated in one-loop approximation using
the fully renormalized hole-GF.

The superconducting temperature $T_{c}$ is calculated from
the linearized form of the Eliashberg equation for the gap-function
\begin{eqnarray}
\phi(k,i\omega_{n})=\sum_{p}\sum_{m}&\lambda&^{12}_{k,k-p}(i\omega_{n}-i\omega_{m})
G_{hh}(p,i\omega_{m})\nonumber \\
\times &G&_{hh}(-p,-i\omega_{m})\phi(p,i\omega_{m}) .
\label{12}
\end{eqnarray}
The first step is a self-consistent calculation of the
normal GF $G_{hh}(k,i\omega_{n})=
(i\omega_{n}+\epsilon_{k}-\Sigma_{hh}(k,i\omega_{n}))^{-1}$
with the self-energy operator (2)
for a given concentration of holes
$
\delta=\dfrac{1}{2}+\dfrac{2T}{N}\sum_{k}^{}\sum_n G_{hh}(k,i\omega_{n}).
$

\begin{figure}
\vskip  0cm
\centerline{\epsfig{file=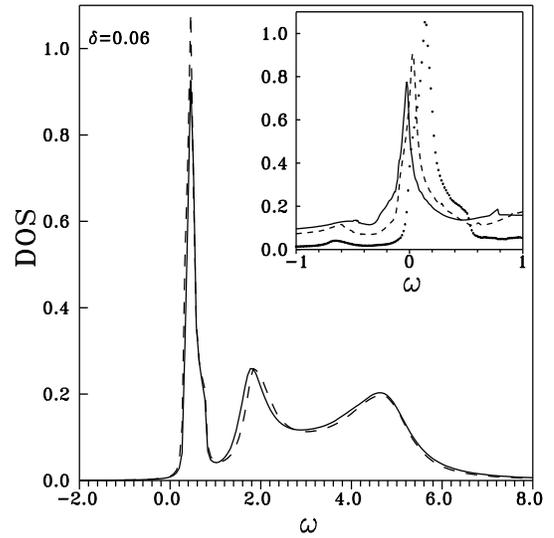,width=7cm,height=7cm}}
\vskip 0.5cm
\caption{The density of states (DOS) for the hole concentration
$\delta=0.06$ at $T=0.012 t$.
The solid (dashed) line  corresponds to calculations with the
full (zero order)  magnon Green function.
In the inset DOS is given for $\delta = 0.1$, 0.25, 0.35
(from right to left) for the zero order magnon  spectra.}
\label{Figone}
\end{figure}

The numerical calculations were performed using
fast Fourier transformation (FFT)~\cite{18}
for a  mesh of 64$\times$64  $k$-points in
the full Brillouin zone  ($0\le k_x, k_y\le 1$), in units of $2\pi$.
In the summation over the  Matsubara frequencies we used up to
200-700 points  with constant cut-off $\omega_{max} = 10t$.
The FFT for the momentum integration is possible due to the particular
momentum dependence of $g(k,q)$.
Usually 10 -- 30 iterations
were needed to obtain a solution for the self energy with an accuracy of order
0.001. Pad\'e approximation was used to calculate the hole spectral function
$A_{}(k,\omega)=-\dfrac{1}{\pi}\mbox{Im }
\langle \langle h_{k}\mid h_{k}^{+}\rangle \rangle _{\omega+i\epsilon}$
and the density of states (DOS) $A_{}(\omega)$
on the real frequency axis.
In Fig. 1 results for $A_{}(\omega)$ of the $t$-$t'$-$J$ model
are shown for various doping
concentrations. The peak in the DOS of width $\Delta W \leq J$ near
the chemical potential $\mu=0$ results from the shallow quasiparticle
dispersion $E(k)$ along the AF-zone boundary (Fig. 2(a)).
We find that the shape of the quasiparticle dispersion  even at
$\delta \sim 0.25$ is still similar to the shape of the dispersion in
the single hole case. Yet a rigid band description fails since the
scale $\Delta W$ and the total quasiparticle bandwidth $W$ grow
significantly with $\delta$. The peak of the DOS coincides with $\mu$
at the crossover from hole to electron like Fermi surfaces (FS). This
occurs at a characteristic concentration which depends on $t'$.
In Fig. 2(b) the FS at $\delta =0.25$ is shown for the two models
studied in this paper, the $t$-$J$ and the
$t$-$t'$-$J$ model with $t'=-0.1 t$. Our unit of energy is $t=1$ (in
reality $t\sim 0.4 eV$ for $CuO_2$ planes) and $J/t=0.4$.
The crossover from hole- to electron-like FS is consistent with the
variation of the Hall constant in $La_xSr_{2-x}CuO_4$.


\begin{figure}
\vskip  0cm
\centerline{\epsfig{file=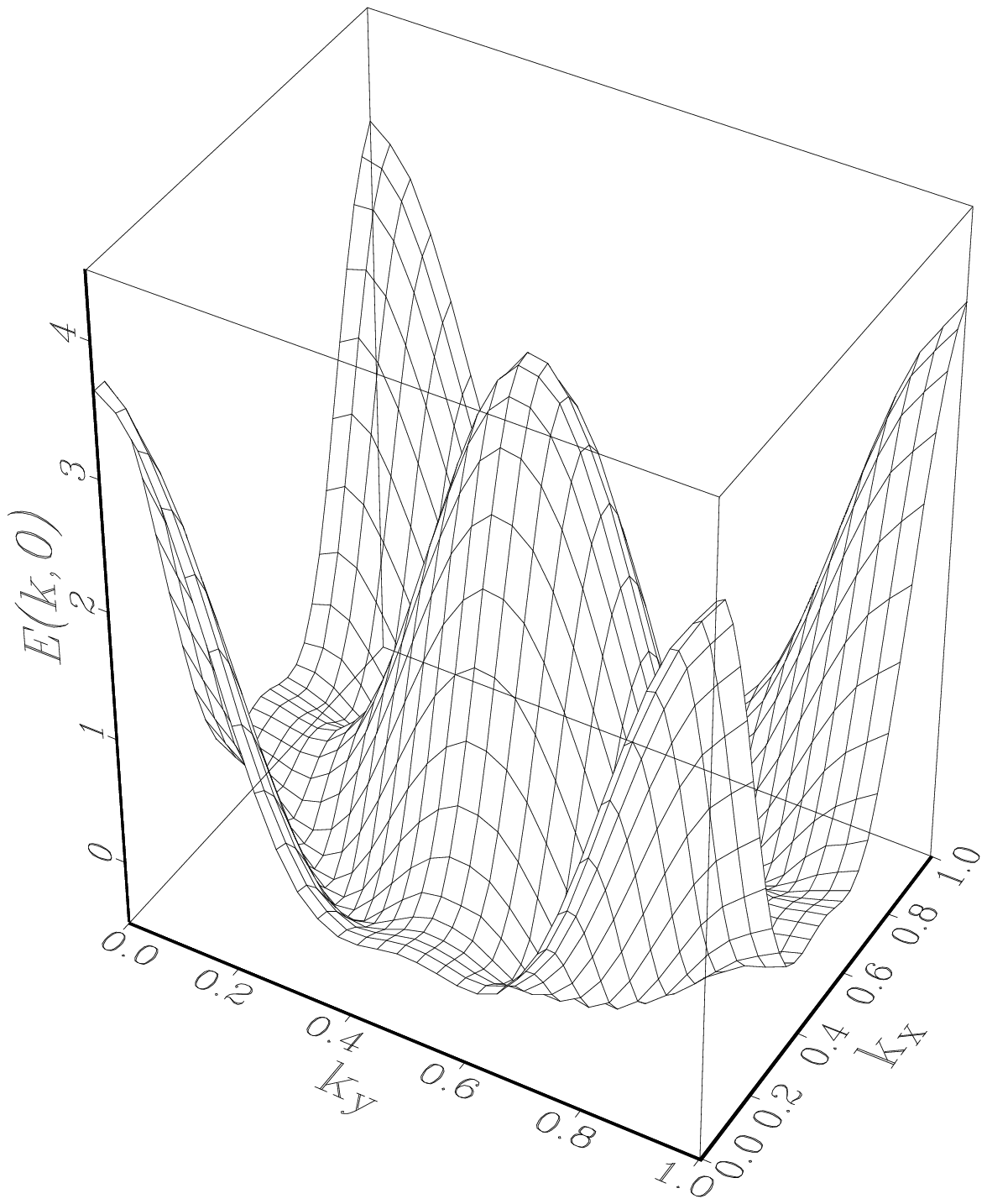,width=6cm,height=6cm}}
\vskip 0.4cm
\centerline{\epsfig{file=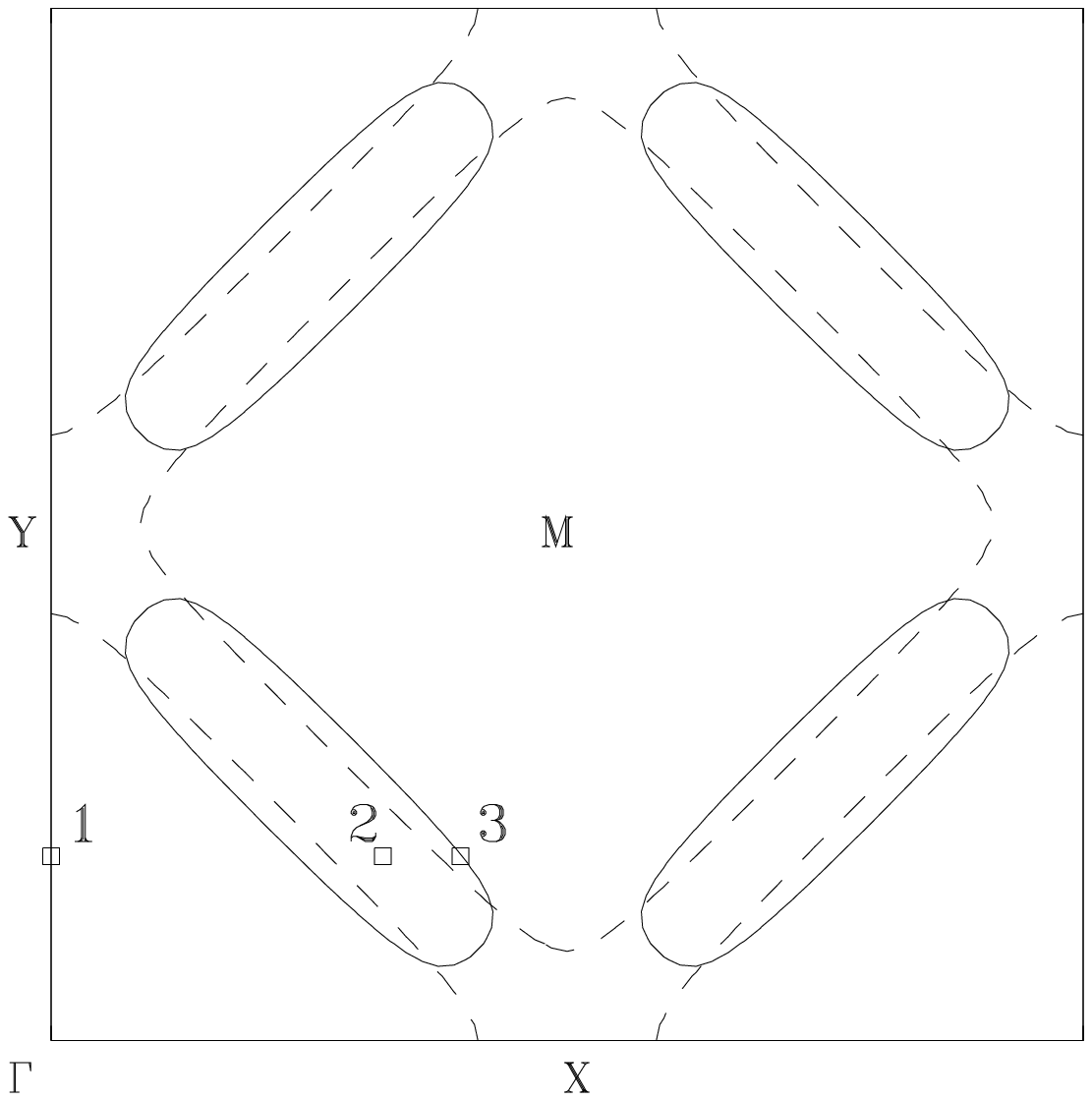,width=5cm,height=5cm}}
\vskip 0.4cm
\caption{(a) The  quasiparticle spectrum $E(\mbox{\bf k})$
and (b)
the Fermi surface (FS) $E(\mbox{\bf k}_{F})=0 $ of the $t-t^{'}-J (t-J)$
model for  $\delta=0.25$ is given by the solid (dashed) line.}
\label{Figtwo}
\end{figure}


The particle-hole renormalization of the magnon propagator $D$ leads
to an instability at small $q$ indicating the disappearance of AF long
range order. In Fig.1 we compare the DOS $A(\omega)$ at $\delta=0.06$
calculated with $D$ and $D_0$, which is not ill behaved. The small-$q$
instability has only small effects on $A(k,\omega)$ and $A(\omega)$
since in the small $q$-regime the spin-charge coupling is
small. Therefore we performed our calculations at higher $\delta$ with
$D_0$. Our main assumption here is, that the spin polaron approach
gives a reliable description also in the spin liquid regime provided
the AF correlation length is sufficiently large compared to the Cooper
pair and polaron radius. The latter quantity is 2 lattice
constants  for $J/t=0.4$\cite{18b} .


The momentum dependence of the gap function
$\Delta(k,\omega=0)$,
$\Delta(k,\omega)=\phi(k,\omega)/Z(k,\omega)$,
is shown in Fig. 3(a)  for  $\delta=0.25$ and
$T/T_{c}\approx 0.8$.
Here $Z(k,\omega)$ is an analytical continuation of the
Eliashberg function
$Z(k,i\omega_n)=(1-Im\Sigma(k,i\omega_n)/\omega_n)^{-1}$.
The gap function  has the  typical $d$-wave symmetry
with two ridges resulting from sharp changes of the interaction function at
the FS. In Fig.~3(b) the frequency dependence of
$Re\Delta(\mbox{k},\omega)$
is shown for a set of $(k_{x}, k_y)$ points marked in
Fig.2b:
(1) inside the FS, (0,~0.19),
(2) at the AF-zone boundary, (0.31,~0.19),
(3) near the FS, (0.38,~0.19).
The gap function changes sign after crossing the $k_x=k_y=0.19$ point
where it is equal to zero.
It is interesting that
the characteristic energy cutoff for the pairing theory,  which is of order
$J\simeq 0.4$  away from the FS (curve 1 ),
becomes much smaller near the FS (curves 2 and 3).
The sharp change of the real part
and the quite large values of $Im\Delta(k,\omega)$ near the FS
differ from the results for conventional superconductors.
Since the Fermi energy $E_F$ is of the order of the exchange energy $J$
{\it all quasiparticles} contribute to the pairing state contrary to
the weak coupling case in conventional superconductors.


\begin{figure}
\vskip  0cm
\centerline{\epsfig{file=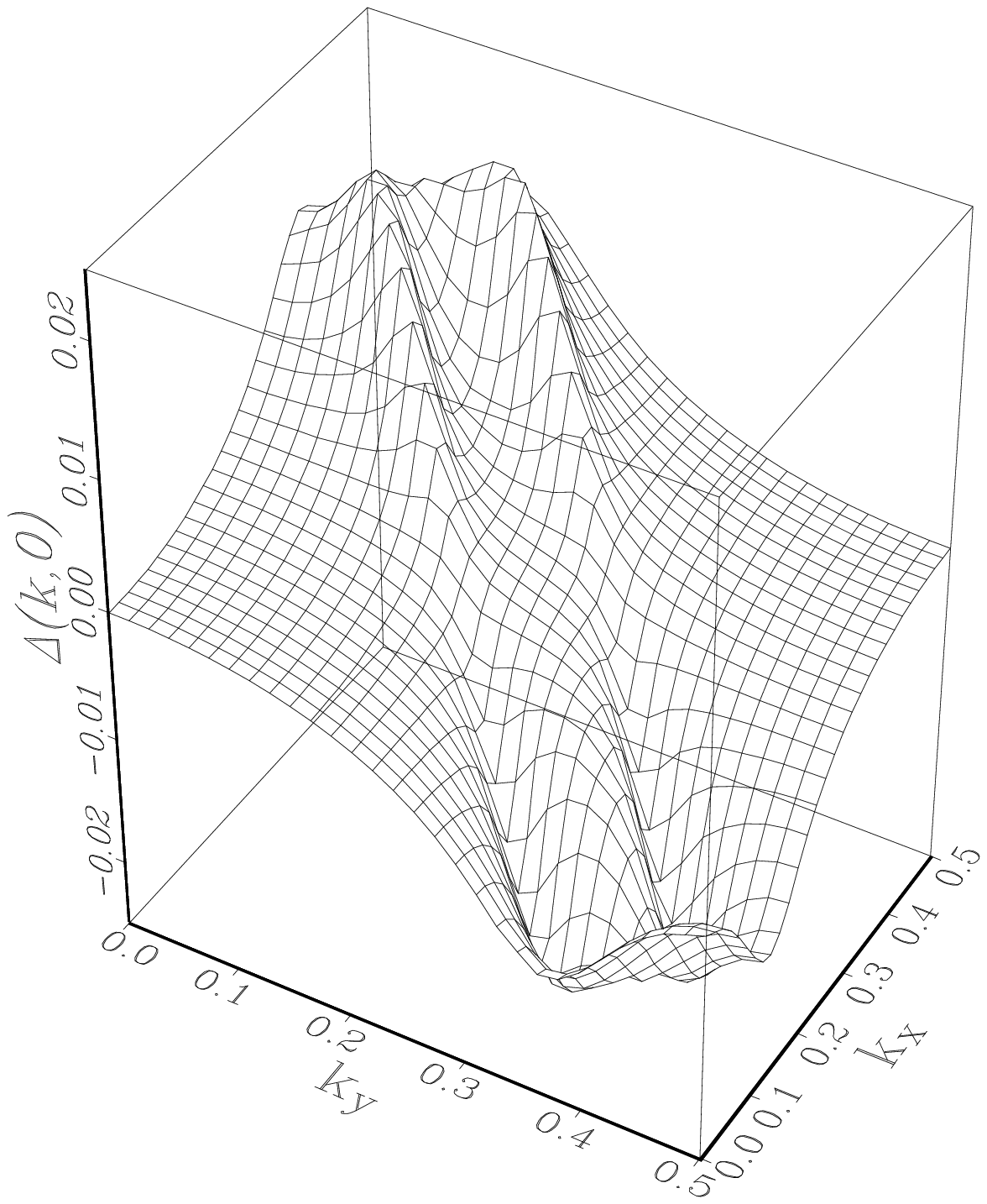,width=6cm,height=6cm}}
\vskip 0.4cm
\centerline{\epsfig{file=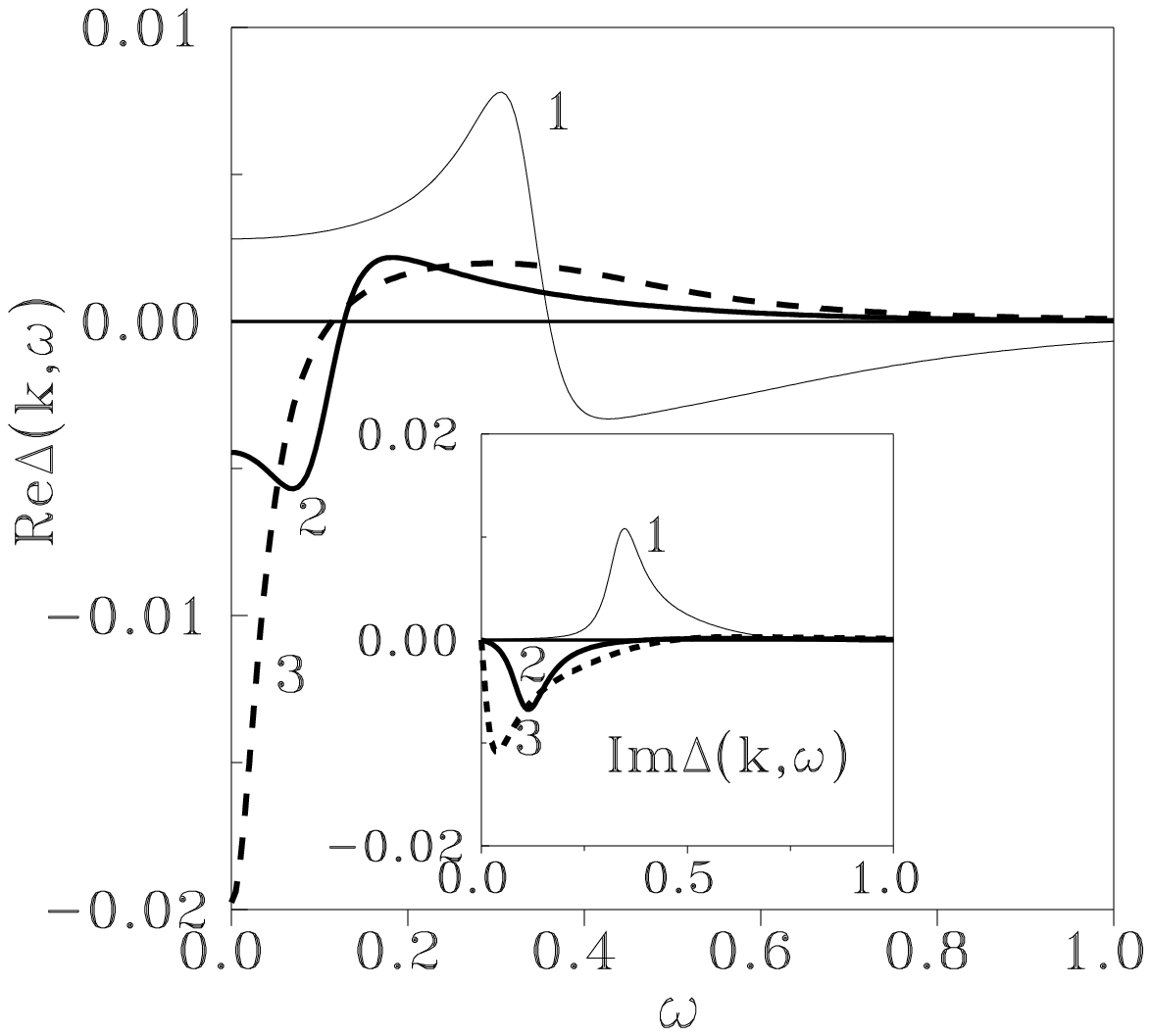,width=6cm,height=6cm}}
\vskip 0.4cm
\caption{(a) The gap function $\Delta(\mbox{\bf k},\omega=0)$
versus $\mbox{\bf k}$ (in units of $2\pi/a$)  and (b)
Re~$\Delta(\mbox{\bf k},\omega)$ (Im~$\Delta(\mbox{\bf k},\omega)$
in the inset) versus $\omega$ for a set of $(k_{x}, k_y)$ points shown
in Fig.2b for $t$-$t^{'}$-$J$ model
 ( $\delta=0.25$ and $T/T_{c}\approx 0.8$.)}
\label{Figthree}
\end{figure}

The transition temperature $T_c$ is determined
as the temperature where the highest eigenvalue of the linearized
Eliashberg equation becomes unity.
In all cases the symmetry of the corresponding eigenfunction
$\phi(k,\omega)$ is $d_{x^2-y^2}$.
In Fig.~4
the dependence of $T_c$ on hole concentration is
shown for $t'=-0.1 t$ and $t'=0$.
These results are quite different from the monotonic increasing  $T_c$
obtained within the weak coupling limit of the BCS equation in ~\cite{8b}
and the maximum of $T_c$ found in ~\cite{1c} near half filling.
In our case the maximum of $T_c$ at $\delta \simeq 0.25$
(or at $\delta \simeq 0.20$ for $t'=0$) results from the Fermi
level crossing of the peak in the density of states which coincides with the
change of the FS topology.


\begin{figure}
\vskip  0cm
\centerline{\epsfig{file=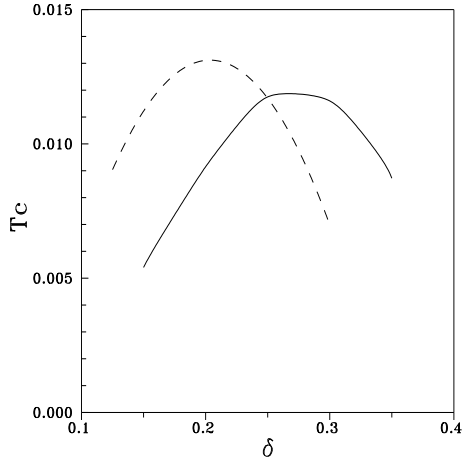,width=6cm,height=6cm}}
\vskip 0.4cm
\caption{The superconducting temperature $T_{c}$ versus
hole concentration $\delta$ for  $J=0.4$, $t'=-0.1t$ (solid line)
and $t'=0$ (dashed line).}
\label{Figfour}
\end{figure}

We have also studied the dependence of $T_{c}$ on the exchange
energy for $J\le 4$. $T_{c}$ increases with $J$ and saturates at
$T_{c}\simeq 0.025 t$ for $J\simeq 3$.
However, we have not obtained a large  drop of $T_{c}$ for $J>3$
observed in small cluster calculations near phase separation ~\cite{1ba},
which is beyond the scope of our study.

In summary, we have solved numerically Eliashberg equations
for the strong coupling spin-polaron model.
We have calculated the quasiparticle spectrum
of spin polarons in the normal state and shown that they
undergo superconducting
$d$-wave pairing mediated by spin fluctuations.
The high values of superconducting temperature  and its doping
dependence $T_{c}(\delta)$ is explained by a large peak in the density of
states of the spin polaron quasiparticles in the vicinity of the
chemical potential.
A key difference from the van Hove scenario, however, is that
the  quasiparticle density  of states has a width given by the
interaction energy $J$ which is similar to  the pairing energy.
We have found unconventional behavior  for
the $d$-wave gap function (a sharp change with energy and large damping
near the FS) which suggests an explanation for some of anomalous
properties of cuprate superconductors observed in tunneling experiments
(v-shape gap and large imaginary part), infrared absorption (no visible
gap or gapless superconductivity), ARPES (a line of gap nodes along
$(\pi,\pi)$ direction ~\cite{19}), etc.
Our calculations are based on a two sublattice representation,
which is suggested to provide a reasonable description of
spin polaron quasiparticles and their pairing even in the spin liquid regime,
i.e. for hole concentrations where the AF correlation length is sufficiently
large.

An important difference between the phenomenological spin-fluctuation
theory and our approach is that pairing is dominated in the former by
$q \sim (\pi,\pi)$ scattering and energy transfers $\Delta E <
50 mev$, whereas in our calculations high-energy spin-fluctuations
with $q $ near the AF-zone boundary are most important.
Higher energy neutron scattering data in this momentum and energy
range is however not yet available.

We thank H.~Eschrig, P.~Fulde and G. Khaliullin  for helpful discussions.
One of the authors (N.P.) thanks Yukawa Institute for Theoretical Physics
for hospitality where
part of the work has been carried out.
We acknowledge the financial support by the Russian State Program
``High-Temperature Superconductivity''  (Grant  92052)
and the support by the
Heisenberg-Landau Program.


\end{multicols}
\end{document}